%% file: main.tex
\documentclass[%
11pt,
onecolumn,
tightenlines,
superscriptaddress,
notitlepage,
preprintnumbers,
nofootinbib,
amsmath,amssymb,amsthm,
aps,
pre,
eqsecnum
]{revtex4-2}

\include{preamble}

\begin{document}

%\title{Symmetry-adapted Markov chain Monte Carlo for modeling the influence of dipole-dipole interactions on electromechanical coupling in polymer chains}
\title{Discrete-symmetry-adapted Markov chain Monte Carlo for the electro-elasticity of polymers: chain taut, collapse, and order}
\author{Matthew Grasinger}
\email{matthew.grasinger.1@us.af.mil}
\affiliation{Foundational Technologies Directorate, Air Force Research Laboratory}

\begin{abstract}
Dielectric elastomers are promising candidates for soft robotics, wearable electronics, and adaptive sensing, but their widespread adoption is hindered by the high electric fields required for significant actuation. Maximizing electromechanical coupling through an understanding of molecular-scale mechanisms is therefore essential. However, theoretical progress on the role of dipole-dipole interactions between monomers has been limited, in part because the resulting energy landscapes, characterized by multiple symmetric wells separated by high barriers, are difficult to sample with standard molecular simulation methods. This work develops a discrete-symmetry-adapted Markov chain Monte Carlo method that exploits the invariance of dipole-dipole interaction energies under simultaneous reflection of neighboring monomer orientations about the plane orthogonal to the applied electric field. Variable sized clusters of neighboring monomers are constructed and reflected collectively, enabling rapid transitions between symmetric energy wells and rendering feasible simulations that are otherwise intractable due to critical slowing down. Application to freely jointed chains with anisotropic monomer polarizability reveals qualitatively distinct electroelastic phenomena depending on the orientation of the monomer dipole relative to its backbone axis. Field-aligning chains exhibit local straightening, electrically induced tautness, and an apparent compressive stiffness, while field-disaligning chains exhibit local folding and electrically induced collapse at fields one to two orders of magnitude lower. Monomer orientational order is quantified across the applied field and susceptibility parameter space, revealing sharp transitions suggestive of underlying phase transitions. For field-disaligning chains, these orientational transitions correspond directly with sharp changes in chain polarization, linking microstructural rearrangement to the macroscopic dielectric response. The symmetry-adapted approach generalizes naturally to other multifunctional polymer systems whose energy landscapes possess analogous discrete symmetries.
\end{abstract}

\maketitle

%% main text
\section{Introduction} \label{sec:intro}
Dielectric elastomers (DEs) undergo large deformations in response to electric fields, making them attractive candidates for soft actuators, stretchable electronics, and adaptive sensors. Their compliance, light weight, and high energy density enable applications ranging from artificial muscles in soft robotics to energy harvesting and human-machine interfaces~\cite{majidi2014soft,bar-cohen2001electroactive,carpi2011electroactive,yu2025prestretch,cooley2023nonlinear,cohen2017enhancing}. However, realizing this potential requires addressing two critical limitations: the high electric fields needed for significant actuation and the risk of electrical breakdown. Maximizing electromechanical coupling -- through combined material design, device design, and novel coupling mechanisms -- is essential for achieving faster, larger-amplitude responses while operating at safer voltages.
Theoretical and modeling advancements can help with overcoming these limitations, as they allow for the discovery of new coupling mechanisms and the design of material architectures with enhanced properties (e.g., ~\cite{grasinger2021flexoelectricity,hoang2025exploring,grasinger2023polymer,mulderrig2025polydisperse,grasinger2020architected,khandagale2025nonlocal,khandagale2024statistical,cooley2023nonlinear,grasinger2021torque,katusele2025exploiting,friedberg2023electroelasticity,zhao2007electromechanical,zhao2007method,lu2020mechanics,hajiesmaili2021dielectric,kang2026demonstration,martinez2021silico,hajiesmaili2022programmed,cohen2018generalized,zurlo2017catastrophic,cooley2023leveraging}).
To name a few, theory and modelling have been used to \begin{inparaenum}[1)] \item suggest the possibility of new actuation modes~\cite{grasinger2021torque,cohen2018generalized}, \item design architected DEs for morphing into targeted shapes~\cite{martinez2021silico,hajiesmaili2022programmed} and enhanced electromechanical couplings~\cite{grasinger2020architected}, \item predict and delay failure~\cite{zhao2007method,zurlo2017catastrophic}, \item suppress and exploit instabilities~\cite{katusele2025exploiting,katusele2025soft,su2019tuning}, and \item leverage dynamics to access giant deformations~\cite{cooley2023leveraging}. \end{inparaenum}

To effectively inform material design, modeling must bridge the gap between macroscopic behavior and molecular-scale physics. While continuum-based approaches are useful for simulating complex geometries and experimental boundary conditions, they often lack the resolution to provide insights based on the detailed molecular and macro-molecular responses. Statistical mechanics can act as a bridge in this regard, accounting for entropy and the mechanisms for electromagnetic interactions (e.g., polarization) that originate at the monomer level~\cite{cohen2016electroelasticity,grasinger2020statistical,grasingerIPforce,grasinger2021torque,grasinger2021flexoelectricity,itskov2018electroelasticity,khandagale2025nonlocal,khandagale2024statistical,friedberg2023electroelasticity}. 
%This multi-scale perspective can aid in understanding how the collective behavior of monomers polymer chains dictates the overall performance of the elastomer.
Following Stockmayer's observation that the dielectric response of polymers is often transversely isotropic, with the monomer axis being the special direction~\cite{stockmayer1967dielectric}, seminal work by Cohen et al.~\cite{cohen2016electroelasticity} incorporated this monomer constitutive response into a statistical mechanics formulation for dielectric polymer chains.
Subsequent work extended this framework to derive approximations in various regimes, including large chain stretch and alternative ensembles~\cite{grasinger2020statistical,grasinger2020architected,grasinger2021torque,grasingerIPforce,itskov2018electroelasticity}.
However, these analytical treatments rely on the assumption that dipole-dipole interactions between monomers are negligible relative to thermal energy and/or field-dipole interaction energy.
This assumption may break down when the dipole susceptibility is large or when chain configurations bring monomers near to each other.

Recent work by Khandagale et al.~\cite{khandagale2024statistical,khandagale2025nonlocal} addressed this using a statistical field theory approach capable of modeling dielectric polymer chains with excluded volume and nonlocal electrostatic interactions. 
This approach revealed new phenomena, notably an electrically driven chain collapse, which is not predicted by non-interacting models. 
However, some open questions remain. 
The field-theoretic formulation models the chain as a continuous space curve rather than as discrete, rigid links of finite length, and relies on a numerical saddle-point approximation whose range of validity if and when fluctuations about the saddle point are large has not been fully characterized. 
Markov chain Monte Carlo (MCMC) offers an alternative computational approach that naturally accommodates discrete monomers with finite degrees of freedom and does not rely on saddle-point or mean-field approximations.

\subsection{Markov chain Monte Carlo}
Markov chain Monte Carlo (MCMC) is a robust and versatile tool for addressing such high-dimensional statistical problems, with vast applications ranging from computational physics~\cite{fall2023optimized,grasingerIPforce,frenkel2001understanding,bartels1998probability,mezei1987adaptive,brooks2011handbook,landau2021guide,tuckerman2010statistical,krauth2006statistical,krauth2021event} to Bayesian inference and uncertainty quantification~\cite{grasinger2016decision,shields2018adaptive,zhu2018bayesian,brooks2011handbook}. By generating a chain of microstates, MCMC allows for the approximation of phase space averages that would otherwise be infeasible due to the ``curse of dimensionality'' inherent in numerical integration. For dielectric polymer chains, MCMC provides a pathway to simulate the response of interacting monomers where exact analytical solutions do not exist.
Let $\mState$ denote a microstate of the ensemble of interest and let $\pSpace$ denote the phase space; that is, the space of all possible microstates.
Then we denote the probability density of a microstate as $\prob{\mState}$ such that $\prob{\mState} \df{\mState}$ is the probability that the system being in $\left[\mState, \mState + \df{\mState}\right] \subset \pSpace$.
We are interested in averages of observables, $\obs$, over phase space which we denote by
\begin{equation} \label{eq:phase-avg}
	\avg{\obs} \coloneqq \int_{\pSpace} \df{x} \; \prob{x} \obs\left(x\right).
\end{equation}
MCMC consists of randomly generating a chain of microstates, $\chain = \set{\mState_1, ..., \mState_{\nStates}}$, and then making the approximation
\begin{equation} \label{eq:expectation-approximation}
	\avg{\obs} \approx \chainAvg{\obs} \coloneqq \frac{1}{\nStates} \sum_{I=1}^{\nStates} \obs\left(\mState_I\right)
\end{equation}
If the chain of states is properly generated, then $\chainAvg{\obs} \rightarrow \avg{\obs}$ as $\nStates \rightarrow \infty$.
Let \begin{equation}\chain_{\mState, \df{\mState}} \coloneqq \set{\mState \in \chain \: : \: \mState \in \left[\mState, \mState + \df{\mState}\right]}. \end{equation}
Then we desire to sample microstates with the Markov chain such that \begin{inparaenum}[1)] \item \begin{equation} \label{eq:convergence-to-distribution} \setSize{\chain_{\mState, \df{\mState}}} / \nStates \approx \prob{\mState} \df{\mState} \end{equation} for all $\mState \in \pSpace$;
	and, further, \item the goal is that the approximation error vanish as quickly as possible. \end{inparaenum}
One way to guarantee `1)' is to ensure that the chain is \emph{Markovian} (i.e., memoryless), ergodic (i.e., can reach all of $\pSpace$ and is acyclic), and satisfies detailed balance~\cite{sethna2021statistical}:
\begin{equation} \label{eq:detailed-balance}
	\prob{\mState} \transProb{\mState}{\otherMState} = \prob{\otherMState} \transProb{\otherMState}{\mState},
\end{equation}
where $\transProb{\mState}{\otherMState}$ is the probability of transitioning from state $\mState$ to $\otherMState$.
How to ensure the approximation error vanishes as quickly as possible is a more subtle detail.

The efficacy of MCMC, however, can be hampered by the presence of energy barriers. In the context of DEs, the interaction between monomers and the electric field creates an energy landscape with multiple local minima. When these ``islands'' of likely states are separated by high energy barriers (e.g., regions where monomers are forced to align against a preferred direction), standard MCMC methods suffer from ``critical slowing down''. In such cases, the chain becomes localized in a single well and fails to mix rapidly enough to provide a converged, accurate approximation of the ensemble average.
This challenge is not unique to Monte Carlo methods; molecular dynamics also struggles with these landscapes, as traversing an energy barrier is a ``rare event'' that requires significant simulation time to resolve.
This barrier-crossing problem is ubiquitous in statistical mechanics. 
Several approaches aim to address this challenge. Umbrella sampling introduces biasing potentials to flatten energy barriers, allowing the chain to traverse between wells more readily. The true ensemble average is recovered by reweighting:
\begin{equation} \label{eq:umb-sampling}
	\modProb{x} \propto \weight{\mState} \exp\left(-\invT \U\left(\mState\right)\right).
\end{equation}
Then the thermodynamic state variables can be obtained by a modified averaging over the MCMC samples:
\begin{equation} \label{eq:umb-avging}
	\avg{\obs} \approx \frac{\modChainAvg{\obs / \weightSym}}{\modChainAvg{1 / \weightSym}}.
\end{equation}
where $\modChainAvg{.}$ denotes the chain average obtained by sampling using the modified probability distribution, $\modProb{.}$.
There are many subtleties related to choosing an appropriate weight function $\weight{x}$ such that the MCMC sampling is ergodic (i.e., does not become trapped in some subset of $\pSpace$) and has a good convergence rate~\cite{frenkel2001understanding,bartels1998probability,mezei1987adaptive,brooks2011handbook,krauth2006statistical}.
By introducing a pseudopotential, Umbrella sampling enhances the rate of (global) mixing at the expense of sampling by importance; in other words, while the pseudopotential allows the chain to traverse energy barriers, it does so by biasing the sampling toward higher energy (and, hence, lower probability density) states.
As a consequence, the chain spends more iterations sampling regions of $\pSpace$ which are less important in the sense that they have smaller contributions to phase averages.
Nonetheless umbrella sampling has proven to be an effective sampling approach for energy landscapes with barriers.

Alternative approaches include replica exchange (parallel tempering)~\cite{lyubartsev1992new,marinari1992simulated}, which simulates multiple copies at different temperatures and exchanges configurations between them, and Hamiltonian Monte Carlo~\cite{beskos2013optimal,betancourt2016identifying}, which uses molecular dynamics trajectories as the proposal moves. However, these methods require extensive parameter tuning and increased computational cost.
This work proposes a fundamentally different approach: rather than fighting the energy barriers or adding computational overhead, we exploit the discrete symmetries inherent in the dielectric polymer energy landscape. By identifying and leveraging group-theoretic structure -- specifically, reflection symmetry about the plane orthogonal to the applied electric field -- we construct variable-sized clusters of monomers that can be simultaneously ``flipped'' between energy wells. This achieves more rapid global mixing without biasing or temperature replicas, providing an efficient and scalable method for sampling complex multiphysics polymer systems.

\subsection{Discrete symmetries, energy barriers, and group theoretic trial moves} \label{sec:discrete-symmetries}

The challenge of barrier-separated energy wells has a classic solution in the Ising model: clustering algorithms that simultaneously flip groups of aligned spins, exploiting the model's reflection symmetry~\cite{wolff1989collective,swendsen1987nonuniversal,sethna2021statistical,fall2023optimized}. 
This circumvents critical slowing down by allowing the chain to jump between symmetric energy wells rather than slowly diffusing through low-probability barrier states.
This principle generalizes beyond the Ising model. Consider an energy landscape with a dominant component that has a discrete symmetry: $\U(\mState) = \U_G(\mState) + \epsilon \U_O(\mState)$, where $\U_G$ is invariant under a group action $\group$ (i.e., $\U_G(\mState) = \U_G(g \cdot \mState)$ for $g \in \group$ and $\epsilon$ is small). When $\group$ is discrete and $\U_G$ has a local minimum $\mStar$, the orbit $\{g \cdot \mStar : g \in \group\}$ defines multiple energy wells separated by barriers. Standard local moves suffer critical slowing down, as the chain becomes trapped in a single well.

Our approach proposes trial moves by: \begin{inparaenum}[1)] \item applying a standard local perturbation and \item applying a randomly sampled group action $g \in \group$. \end{inparaenum}
This enables the chain to jump between symmetric wells efficiently. For dielectric polymer chains in electric fields, the relevant symmetry is
reflection about the plane perpendicular to the field-monomers aligned parallel vs. antiparallel to the field occupy distinct but symmetric energy wells.

\paragraph*{Group theoretic acceleration in particle-scale methods.}
This work is inspired, in part, by the successes of various works that leverage group theory to improve the efficiency of modeling bulk materials, nanostructures, viruses, etc., using molecular dynamics~\cite{james2006objective,dayal2010nonequilibrium} and the Boltzmann equation~\cite{dayal2010nonequilibrium}.
Similar considerations of geometry and symmetry transformations have also been leveraged to develop various clustering-type MCMC algorithms~\cite{dress1995cluster,heringa1998geometric,liu2005generalized,ruuvzivcka2014collective}.
An aim of this work is to explore geometry and symmetry in the context of multiphysics of polymer chains with an emphasis on the importance of discrete symmetries.

\section{Statistical mechanics formulation}
\begin{figure}
	\centering
	\includegraphics[width=0.49\linewidth]{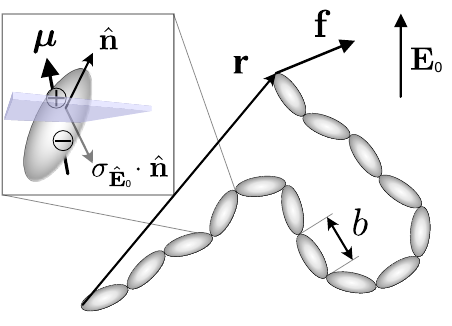}
	\caption{
		\captitle{Freely jointed dielectric polymer chain} with end-to-end vector, $\rvec$, applied force, $\fvec$, and applied electric field, $\efield$ (adapted with permission from~\cite{grasingerIPforce}).
		Each monomer forms an electric dipole in response to the field and has a corresponding interaction energy.
		The monomer-field interaction energy has a discrete symmetry associated with the action $\groupAct{\reflAbout{\edir}}{\nvec}$ where $\nvec$ is the monomer direction (shown in upper left panel).
	}
	\label{fig:de-poly-chain}
\end{figure}

A linear polymer chain is a macromolecule consisting of repeating units (monomers) bonded end-to-end.
(An example polymer chain is shown in \fref{fig:de-poly-chain}.)
We model the chain electromechanics using the freely jointed chain (FJC) approximation~\cite{treloar1975physics}: monomers are rigid rods of length $\mlen$ (the K\"uhn length), free to rotate about bonds, with excluded volume effects neglected.
For a chain of $\numMonomers$ monomers, the maximum end-to-end distance is $\numMonomers \mlen$.
Each monomer's orientation is described by a unit vector, $\nvec$, along its axis, pointing from one bond to the next.
A microstate of the chain is then described by $\mStateP = \left(\nvec_1, \dots, \nvec_{\numMonomers}\right)$~\footnote{
	Note that, when using spherical coordinates to describe $\left(\mathbb{S}^2\right)^\numMonomers$, the unnormalized probability of a microstate must also include the Jacobian: \begin{equation*}
		\prob{\mStateP} \propto \exp\left(-\invT \U\left(\mStateP\right)\right) \prod_{i=1}^{n} \sin \theta_i,
	\end{equation*} where $\theta_i$ is the polar angle of $\nvec_i$.
}.
Here also, $\pSpace = \mathbb{S}^2 \times \dots \times \mathbb{S}^2 = \left(\mathbb{S}^2\right)^\numMonomers$, and
\begin{equation}
	\rvec = \mlen \sum_{i=1}^{\numMonomers} \nvec_i.
\end{equation}

In the presence of an electric field, bound charges on a monomer can separate to form an electric dipole, $\dipoleVec$, where the resistance to charge separation depends on the direction of the field relative to the monomer axis.
We assume the standard anisotropic form \cite{stockmayer1967dielectric,cohen2016electroelasticity}:
\begin{equation} \label{eq:dipole-response}
	%\begin{split}
	%	\dipoleVec\left(\nvec, \efield\right) &= \vperm \dipoleSusceptibility \efield \\ &= \vperm \left[\susPara \nvec \otimes \nvec + \susPerp\left(\idenMat - \nvec \otimes \nvec\right)\right] \efield 
	%\end{split}
	\dipoleVec\left(\nvec, \efield\right) = \vperm \dipoleSusceptibility \efield = \vperm \left(\susPara \nvec \otimes \nvec + \susPerp\left(\idenMat - \nvec \otimes \nvec\right)\right) \efield 
\end{equation}
where $\dipoleSusceptibility$ is the dipole susceptibility tensor, $\susPara$ and $\susPerp$ are the dipole susceptibility along $\nvec$ and the susceptibility in plane orthogonal to $\nvec$, respectively, $\efield$ is the local electric field, and $\vperm$ is the vacuum permittivity.
We call a polymer chain ``field-aligning'' (FA) when $\susPara > \susPerp$ and ``field-disaligning'' (FD) when $\susPara < \susPerp$.
The electrostatic interaction of a single monomer with $\efield$ has two contributions: the energy associated with separating charges and the electric potential of a dipole in an electric field~\cite{grasinger2020statistical}~\footnote{Where $\dipoleSusceptibility^{-1}$ denotes the generalized inverse of $\dipoleSusceptibility$.}:
\begin{equation}  \label{eq:monomer-energy}
	\um = \frac{1}{2\vperm} \dipoleVec \cdot \dipoleSusceptibility^{-1} \dipoleVec - \dipoleVec \cdot \efield = -\frac{1}{2} \dipoleVec \cdot \efield = \frac{\vperm \dsus}{2} \left(\efield \cdot \nvec\right)^2 - \frac{\vperm \susPerp}{2} \emag^2,
\end{equation}
where $\dsus = \susPerp - \susPara$.

A constant force is applied to the end of the chain. %~\footnote{While, for simplicity, only the fixed force ensemble is considered, the principles developed herein would naturally extend to techniques used for efficiently sampling the fixed end-to-end vector ensemble such as the configurational-bias Monte Carlo~\cite{rosenbluth1955monte,frenkel2001understanding}.}.
Given the models for the mechanical and dielectric responses of the monomers in the chain, we arrive at the energy for a dielectric polymer chain:
\begin{equation} \label{eq:U-DE}
	%\U = \sum_{i=1}^{\numMonomers} \left(-\frac{1}{2} \dipoleVec_i \cdot \efield\right) - \fvec \cdot \left(\mlen \sum_{i=1}^{\numMonomers} \nvec_i\right) - \sum_{i=1}^{\numMonomers} \sum_{j=i+1}^{\numMonomers} \frac{3 \left(\dipoleVec_i \cdot \hat{\mathbf{z}}_{ij}\right) \left(\dipoleVec_j \cdot \hat{\mathbf{z}}_{ij}\right) - \dipoleVec_i \cdot \dipoleVec_j}{4 \pi \epsilon_0 z_{ij}^3}
	\U = \sum_{i=1}^{\numMonomers} \left(-\frac{1}{2} \dipoleVec_i \cdot \efield\right) - \fvec \cdot \rvec - \sum_{i=1}^{\numMonomers-1} \left(\frac{3 \left(\dipoleVec_i \cdot \relDirVec{i,i+1}\right) \left(\dipoleVec_{i+1} \cdot \relDirVec{i,i+1}\right) - \dipoleVec_i \cdot \dipoleVec_{i+1}}{4 \pi \epsilon_0 \relPosMag{i,i+1}^3}\right)
\end{equation}
where, $\relPosVec{i,j} = \posVec{i} - \posVec{j}$, $\relPosMag{i,j} = \left|\relPosVec{i,j}\right|$, $\relDirVec{i,j} = \relPosVec{i,j} / \relPosMag{i,j}$,
\begin{equation} \label{eq:monomer-positions}
	\posVec{k} = \posVec{k-1} + \frac{\mlen}{2} \left(\nvec_{k-1} + \nvec_k\right) = \mlen \sum_{i=1}^{k-1} \nvec_i + \frac{\mlen}{2} \nvec_k; \text{ for } k = 2, \hdots, \numMonomers, \qquad \posVec{1} = \frac{\mlen}{2} \nvec_1
\end{equation}
are the monomer positions, and $\fvec$ is the applied force on the chain.
The first term in \eqref{eq:U-DE} corresponds to the electrostatic interaction of the individual monomers with the applied field; the second term represents the work of the force done on the polymer chain; and the last term corresponds to the monomer-monomer interactions of induced dipoles within the monomers of the chain.
For simplicity, the interaction energy only includes Ising-type nearest neighbor terms.
%See~\citet{grasinger2020statistical,grasinger2020architected,grasingerIPforce} for a more in-depth development of \eqref{eq:dipole-response}-\eqref{eq:monomer-positions}.
\Fref{fig:de-poly-chain} shows a schematic of the system (adapted from~\cite{grasingerIPforce}).

\section{Discrete-symmetry-adaptive MCMC}
\subsection{Discrete symmetries and energy barriers}
Let $\edir = \efield / \emag$, $\rotate_{\edir}\left(\varphi\right)$ denote a rotation about $\edir$ by angle $\varphi$, and let $\reflAbout{\edir} = \idenMat - 2 \edir \otimes \edir$ denote a reflection about the plane orthogonal to $\edir$ where $\otimes$ denotes the tensor product and $\idenMat$ is the identity tensor.
It is easy to verify that \begin{inparaenum}[1)] \item $\um\left(\rotate_{\edir}\left(\varphi\right) \nvec\right) = \um\left(\nvec\right)$ for all $\varphi$ and $\nvec$, and  \item $\um\left(\reflAbout{\edir} \nvec\right) = \um\left(\nvec\right)$ for all $\nvec$. \end{inparaenum}
The symmetry of `1)' is a continuous symmetry and does not have any energy barriers associated with it; therefore standard MCMC trial moves (e.g. standard jumps, hybrid Monte Carlo, etc.) are sufficient for sampling this symmetry.
However, `2)', is a discrete symmetry.
For dielectric polymer chains where $\susPara > \susPerp$, $\edir \cdot \nvec = 0$ is an energy barrier which separates wells at $\edir \cdot \nvec = \pm 1$.
When $\invT \vperm \emag^2 \susPara / 2$ is large enough, standard MCMC methods suffer from a poor convergence rate because each monomer in the chain becomes localized at one of the two wells and cannot readily pass through $\edir \cdot \nvec = 0$ to the other well~\cite{grasingerIPforce}.

Given the results of previous sections, we may consider letting $\otherGroup = \set{\iden, \reflAbout{\edir}}$ and augmenting the standard MCMC trial move by sampling and applying an action from $\otherGroup$.
To be precise, we may randomly choose some $i \in \set{1, \dots, \numMonomers}$, perturb $\nvec_i$, and then $\nvec_i \rightarrow \groupAct{g_j}{\nvec_i}$ where $g_j \in \otherGroup$ are sampled via the weights $\groupWeight_1 = \groupWeight_2 = 1/2$.
This sampling algorithm was shown to improve convergence in~\cite{grasingerIPforce} for dielectric chains in which monomer-monomer interactions are negligible.
However, there are additional considerations related to monomer-monomer interactions.

Consider the monomer-monomer interaction term in \eqref{eq:U-DE}.
First, isolate a single contribution to the sum: the interaction between neighbors $i$ and $i+1$.
This interaction term is invariant with respect to simultaneous reflection of monomers $i$ and $i+1$; that is, the action $\nvec_i \rightarrow \reflAbout{\edir} \nvec_i$ and $\nvec_{i+1} \rightarrow \reflAbout{\edir} \nvec_{i+1}$.
Upon application of the action $\nvec_i \rightarrow \reflAbout{\edir} \nvec_i$, the dipole at monomer $i$ takes the form (see \eqref{eq:dipole-response}):
\begin{equation} \label{eq:dipoleAction}
	%\begin{split}
	%	\dipoleVec_i &\rightarrow \vperm \dsus \left(\efield \cdot \reflAbout{\edir}{\nvec_i}\right) \reflAbout{\edir} \nvec_i + \vperm \susPerp \efield \\
	%	&= -\vperm \dsus \left(\efield \cdot \nvec_i\right) \reflAbout{\edir} \nvec_i - \vperm \susPerp \reflAbout{\edir} \efield \\
	%	&= -\reflAbout{\edir} \dipoleVec_i
	%\end{split}
	\dipoleVec_i \rightarrow \vperm \dsus \left(\efield \cdot \reflAbout{\edir}{\nvec_i}\right) \reflAbout{\edir} \nvec_i + \vperm \susPerp \efield = -\reflAbout{\edir} \dipoleVec_i
\end{equation}
such that, when $\nvec_i \rightarrow \reflAbout{\edir} \nvec_i$ and $\nvec_{i+1} \rightarrow \reflAbout{\edir} \nvec_{i+1}$ are applied simultaneously, the inner product $\dipoleVec_i \cdot \dipoleVec_{i+1}$ remains unchanged.
Similarly, by \eqref{eq:monomer-positions}
\begin{equation}
	\relPosVec{i,i+1} = \frac{\mlen}{2}\left(\nvec_i + \nvec_{i+1}\right) \rightarrow \frac{\mlen}{2}\left(\reflAbout{\edir}\nvec_i + \reflAbout{\edir}\nvec_{i+1}\right) = \reflAbout{\edir} \relPosVec{i,i+1}
\end{equation}
upon simultaneous reflection of $i$ and $i+1$.
Both quantities: $\relPosMag{i,i+1}$ and
\begin{equation}
	\left(\dipoleVec_i \cdot \relDirVec{i,i+1}\right) \left(\dipoleVec_{i+1} \cdot \relDirVec{i,i+1}\right)
\end{equation}
are invariant with respect to the simultaneous reflection of monomers $i$ and $i+1$ (i.e., $\nvec_i \rightarrow \reflAbout{\edir} \nvec_i$ and $\nvec_{i+1} \rightarrow \reflAbout{\edir} \nvec_{i+1}$). The interaction energy is therefore also invariant.

\paragraph*{Remarks.} We make (and will later refer) to the following remarks:
\begin{enumerate}
	\item While the interaction energy between monomers $i$ and $i+1$ is invariant with respect to simultaneous reflection, the interaction energy can vary greatly when a monomer is reflected but one of its neighbors is not.
	\label{rem:neighbor-interaction-1}
	\item 
    Recall, a polymer chain is ``field-aligning'' (FA) when $\susPara > \susPerp$ and ``field-disaligning'' (FD) when $\susPara < \susPerp$.
    For FA chains, aligned neighbors (i.e., $\nvec_i = \nvec_{i+1}$) are always energetically favorable. 
    Nearly anti-aligned neighbors ($\nvec_i \rightarrow -\nvec_{i+1}$) cause dipoles that are aligned, side-by-side, and with separation that is vanishing; thus, the interaction energy diverges.
    For FD chains, the situation is more nuanced. When neighboring monomers are confined to the plane transverse to the applied field, aligned neighbors are again favorable -- not because the dipole geometry is optimal (the dipoles are in fact side-by-side and co-aligned with the field), but because the separation remains maximal and the divergent anti-aligned configuration is avoided. However, when neighboring monomers are folded such that they are nearly stacked along the field axis, locally folded (anti-aligned) configurations can instead be favorable, as they arrange the FD induced dipoles approximately end-to-end along the field direction.
	\label{rem:neighbor-interaction-2}
	\item A straightforward way of decomposing $\U$ into $\UGroup + \smallParam \UO$ would be
	\begin{equation}
		\begin{split}
			\UGroup &\coloneqq \sum_{i=1}^{\numMonomers} \left(-\frac{1}{2} \dipoleVec_i \cdot \efield\right) - \sum_{i=1}^{\numMonomers-1} \left(\frac{3 \left(\dipoleVec_i \cdot \relDirVec{i,i+1}\right) \left(\dipoleVec_{i+1} \cdot \relDirVec{i,i+1}\right) - \dipoleVec_i \cdot \dipoleVec_{i+1}}{4 \pi \epsilon_0 \relPosMag{i,i+1}^3}\right) \\
			\smallParam \UO &\coloneqq -\fvec \cdot \rvec
		\end{split}.
	\end{equation}
	Here $\group = \set{\iden, \reflAll}$ where $\groupAct{\reflAll}{\mStateP} = \left(\reflAbout{\edir} \nvec_1, \dots, \reflAbout{\edir} \nvec_{\numMonomers}\right)$.
	The drawbacks with taking $\otherGroup = \group$ are \begin{inparaenum}[1)] \item application of $\reflAll$ causes a significant change in $\rvec$ as $\rvec \rightarrow \reflAbout{\edir} \rvec$, and, consequently, a significant change in the work of the applied force; and \item the orbit of $\genBy{\otherGroup}$ will not contain many additional microstates in $\pSpace$ (it contains at most two distinct elements). \end{inparaenum}
	\label{rem:refl-chain}
\end{enumerate}
By \ref{rem:neighbor-interaction-1}, it is, in some way, suboptimal to merely augment the standard MCMC jumps by reflecting a single monomer about the plane orthogonal to $\edir$.
By \ref{rem:refl-chain}, it is also suboptimal to merely augment the standard jumps by simultaneously reflecting all of the monomers in the chain.
This leads to the idea of clustering (e.g., ~\cite{wolff1989collective,swendsen1987nonuniversal}, \S 5.2.3 of~\cite{krauth2006statistical}, and \S 14 of~\cite{frenkel2001understanding}).

\subsection{Symmetry-adaptive clustering}
The key idea of the discrete-symmetry-adapted MCMC is to randomly construct a variable sized cluster of neighboring monomers and reflect them simultaneously.
Because clusters are (generally) larger than a single monomer, this transitions the microstate between disparate energy wells.
Because the clusters are of variable size, successive action samples a greater number of distinct microstates.
Both properties enhance the overall mixing of the chain.

Randomly choose some monomer, $k$.
Then we add monomer $k+1$ to the cluster with probability, $\clusterProbSym_{k,k+1}$.
If monomer $k+1$ was added to the cluster, we continue by adding its next neighbor $k+2$, and so on, again with probability $\clusterProbSym_{k+1,k+2}$, until either the trial expansion of the cluster has been rejected or we have reached the end of the chain.
Upon completion, let the cluster expand in the opposite direction, starting with $k-1$, then $k-2$, and so on, until either a trial expansion has been rejected or we have reached the beginning of the chain.
Then sample $\otherGroup = \set{\iden, \reflAbout{\edir}}$ uniformly (i.e., each element with a probability $1/2$) and apply its action to each monomer in the cluster.
%(Of course, depending on the problem, $\otherGroup$ and its weights can be more general than this.)

One must ensure the probabilities are formulated, and the clustering process formalized, in ways that guarantee the Markov chain converge to the underlying distribution; recall the sufficient condition for this is \emph{detailed balance} (equation \eqref{eq:detailed-balance}).
We decompose each step in the Markov chain into two parts: \begin{inparaenum}[1)] \item proposing a trial move and \item accepting or rejecting the trial move. \end{inparaenum}
Let $\trialProb{\mStateP}{\otherMStateP}$ denote the probability of proposing a trial move from state $\mStateP$ to $\otherMStateP$ and $\accProb{\mStateP}{\otherMStateP}$ denote the probability of accepting the trial move $\mStateP \rightarrow \otherMStateP$.
We prescribe the trial and acceptance move probabilities such that they are independent:
\begin{equation}
	\transProb{\mStateP}{\otherMStateP} = \trialProb{\mStateP}{\otherMStateP} \accProb{\mStateP}{\otherMStateP}.
\end{equation}
Then detailed balance can be satisfied by the Metropolis-Hasting acceptance criteria:
\begin{equation} \label{eq:MH-critera}
	\accProb{\mStateP}{\otherMStateP} = \min \set{1, \left(\frac{\prob{\otherMStateP}}{\prob{\mStateP}}\right) \underbrace{\left(\frac{\trialProb{\otherMStateP}{\mStateP}}{\trialProb{\mStateP}{\otherMStateP}}\right)}_{= \hastings{\otherMStateP}{\mStateP}} },
\end{equation}
where $\hastings{\otherMStateP}{\mStateP}$ is the Hastings factor.

Next we specify the cluster expansion probability, $\clusterProbSym_{(.,.)}$, and determine the Hastings factor, $\hastings{\otherMStateP}{\mStateP}$.
For simplicity, we formulate the $\clusterProbSym_{i,j}$ such that it is symmetric with respect to transposition of its indices (i.e., $\clusterProbSym_{i,j} = \clusterProbSym_{j,i}$).
If the cluster spans from monomer $r$ to monomer $s$, then
\begin{equation} \label{eq:cluster-prob}
	\trialProb{\mStateP}{\otherMStateP} = \frac{1}{2} \tIn \left(1 - \clusterProbSym_{r-1,r}\right) \clusterProbSym_{r,r+1} \clusterProbSym_{r+1,r+2} \dots \clusterProbSym_{s-1,s} \left(1 - \clusterProbSym_{s,s+1}\right),
\end{equation}
where $1/2$ is the probability of sampling either $\iden$ or $\reflAbout{\edir}$ from $\otherGroup$, $\tIn$ is the probability of beginning the construction of the cluster by randomly choosing some monomer within it, $\left(1 - \clusterProbSym_{r-1,r}\right)$ is the probability of not adding monomer $r-1$ to the cluster, $\clusterProbSym_{r,r+1}$ is the probability of expanding the cluster from $r$ to $r+1$ (or vice versa), etc.~\footnote{
	Here the condition $\clusterProbSym_{i,j} = \clusterProbSym_{j,i}$ simplifies the calculation considerably.
	Because the probability is the same whether the cluster is expanding in increasing index (i.e. from $r$ to $r+1$) or decreasing index (i.e. from $r+1$ to $r$), we do not need to explicitly consider the monomer at which the cluster started.
}
The probability $\trialProb{\mStateP}{\otherMStateP}$ is the product over the probabilities of each of the individual cluster expansion events that make up the trial proposition because they are mutually independent.
Given \eqref{eq:cluster-prob}, the Hastings factor is
\begin{equation} \label{eq:cluster-prob-1}
	\hastings{\otherMStateP}{\mStateP} = \frac{\left(1 - \clusterProbSym'_{m-1,m}\right) \clusterProbSym'_{m,m+1} \clusterProbSym'_{m+1,m+2} \dots \clusterProbSym'_{n-1,n} \left(1 - \clusterProbSym'_{n,n+1}\right)}{\left(1 - \clusterProbSym_{m-1,m}\right) \clusterProbSym_{m,m+1} \clusterProbSym_{m+1,m+2} \dots \clusterProbSym_{n-1,n} \left(1 - \clusterProbSym_{n,n+1}\right)},
\end{equation}
where $\clusterProbSym_{i,j}$ and $\clusterProbSym'_{i,j}$ are the cluster expansion probabilities for the forward and reverse trial moves, respectively (i.e., $\mStateP \rightarrow \otherMStateP$ and $\otherMStateP \rightarrow \mStateP$, respectively); and where the $\tIn$ factors cancel because $\tIn$ is independent of the current state and only depends on the cluster size (which is the same for both the forward and reverse moves).
For convenience and efficiency of mixing, it is often desirable to keep the Hastings factor near to $1$ when possible.
This motivates choosing probabilities such that $\clusterProbSym_{i,j} = \clusterProbSym'_{i,j}$ when possible.
A general form of this is to let $\clusterProbSym_{i,j}$ be a function of the state variables of the polymer chain and other properties of the thermodynamic system which are invariant with respect to the action $\nvec_i \rightarrow \reflAbout{\edir} \nvec_i$ and $\nvec_{j} \rightarrow \reflAbout{\edir} \nvec_{j}$:
\begin{equation} \label{eq:cluster-prob-func}
	\clusterProbSym_{i,j} = \clusterProb{\nvec_i \cdot \nvec_j, \left(\nvec_i \cdot \edir\right)^2, \left(\nvec_j \cdot \edir\right)^2, \dots; \invT, \efield, \susPara, \susPerp, \mlen, \numMonomers, \dots}.
\end{equation}
In this case, \eqref{eq:cluster-prob-1} simplifies to
\begin{equation} \label{eq:cluster-prob-2}
	\hastings{\otherMStateP}{\mStateP} = \frac{\left(1 - \clusterProbSym'_{m-1,m}\right) \left(1 - \clusterProbSym'_{n,n+1}\right)}{\left(1 - \clusterProbSym_{m-1,m}\right) \left(1 - \clusterProbSym_{n,n+1}\right)},
\end{equation}
Here, motivated by \fref{rem:neighbor-interaction-1} and \ref{rem:neighbor-interaction-2}, the simple choice of
\begin{equation} \label{eq:cluster-prob-func-final}
	\clusterProbSym_{i,j} = \begin{cases}
		\clusterProb{\nvec_i \cdot \nvec_j} = \frac{1}{2}\left(1 + \sign\left(\dsus\right) \nvec_i \cdot \nvec_j\right) & \left|i - j\right| = 1 \\
		0 & \text{otherwise}
	\end{cases}
\end{equation}
is made, where $\sign\left(\dsus\right)$ is the sign of $\dsus$, as it has the property that, if $\nvec_i = \nvec_j$ and $\susPara > \susPerp$, which is energetically favorable, monomer $i$ and $j$ are always clustered together and, consequently, will continue to be aligned after the reflection action on the cluster.
Whereas, when $\nvec_i = -\nvec_j$, which is energetically unfavorable, monomer $i$ and $j$ are never clustered together and, consequently, $\nvec_i \neq -\nvec_j$ after the reflection action on the cluster (except in the special case of $\nvec_c \cdot \edir = 0$ where $c$ equals whichever of the two monomers, $i$ or $j$, is contained in the cluster).
It is possible that more sophisticated forms of \eqref{eq:cluster-prob-func} (e.g., including more information such as temperature, electric field, etc., or considering the relative orientations of monomers that are near to each other in space due to chain folding, but topologically distant along the chain backbone) could be constructed with better convergence properties than \eqref{eq:cluster-prob-func-final}; this is a potential topic for future work. %~\footnote{
	%	In principle, any function of the form $\text{const.} \times \left(1 - \sgn\left(\nvec_i \cdot \nvec_j\right) \left|\nvec_i \cdot \nvec_j\right|^{\eta}\right)$ would have similar limiting behavior.
	%}.
However, it will be shown that for the examples considered herein, \eqref{eq:cluster-prob-func-final} performs well.

\section{Results}
\subsection{Verification of symmetry-adapted trial moves}
The symmetry-adapted clustering algorithm for dielectric polymers is verified here.
For the special case of when \begin{inparaenum}[1)] \item monomer-monomer interactions are negligible, and \item the force is aligned or antialigned with the field direction (i.e., $\fvec / \fmag = \pm \edir$), an exact solution to the statistical mechanics formulation is possible (see~\cite{grasingerIPforce} \S 5.3 for more detail). \end{inparaenum}
To verify the proposed approach, the interaction term in the MCMC simulations is (temporarily) dropped, and its results are compared with the exact solution.
Let $\cStretch = \rmag / \numMonomers \mlen$ denote the (absolute) stretch of the polymer chain.
Then the response $\cStretch = \cStretch\left(\fvec, \dots\right)$ also depends on the temperature, electric field, dielectric properties, and other state variables of the polymer chain.
Consider the case of $\susPara > 0$ and $\susPerp = 0$ since the energy barriers, induced discrete symmetries, only occur when $\susPara > \susPerp$.
Then it is energetically favorable for monomers to align with the axis of the electric field (i.e. $\nvec = \pm \edir$).
As a result, the polymer chains become less stiff when stretched in $\pm \edir$~\cite{grasingerIPforce,grasinger2020statistical,grasinger2020architected,grasinger2021torque,cohen2016electroelasticity,cohen2016electromechanical}.

\paragraph*{MCMC parameters.}
Temperature and dielectric properties are constant at $\invT = 1$, $\vperm = 1$, and $\susPara = 1$, respectively, and $\numMonomers = 100$.
Monomer orientations were parameterized by $\theta_i$ and $\phi_i$, the polar angle and azimuth angle of monomer $i$, respectively.
The standard trial moves perturbed the angle of a randomly chosen monomer, $k$: $\theta_k \rightarrow \theta_k + \Delta \theta_k$ and $\phi_k \rightarrow \phi_k + \Delta \phi_k$ where $\Delta \theta_k$ and $\Delta \phi_k$, were sampled uniformly on $\left[-\delta \theta, \delta \theta\right]$ and $\left[-\delta \phi, \delta \phi\right]$, with initial maximum step sizes $\delta \theta = 3 \pi / 16$ and $\delta \phi = 3 \pi / 8$

\paragraph*{Adaptive sampling.}
The step sizes were adjusted every $2500$ steps to maintain an acceptance ratio between $0.15$ and $0.4$: if the acceptance ratio exceeded $0.4$, step sizes increased by $\delta \Box \rightarrow \xi \delta \Box$; if the acceptance ratio fell below $0.15$, step sizes decreased by $\delta \Box \rightarrow \delta \Box / \xi$, where $\xi = 1.1$.

\paragraph*{Burn-in and replicas.}
A burn-in schedule (similar to simulated annealing) started the temperature high and gradually reduced to the target value through $\invT = \left\{10^{-3}, 10^{-2}, 10^{-1}, 0.5, 1\right\}$. 
For each burn-in temperature, $50000$ steps were completed.
Each case of system parameters was run with and averaged over $25$ replicas (i.e., $25$ independent copies of the system) in order to further mitigate issues of MCMC localization.
All simulations were run for an additional $10^6$ steps after burn-in (total: $25 \times 10^6$ samples per parameter set).

\begin{figure*}
	\centering
	\includegraphics[width=\linewidth]{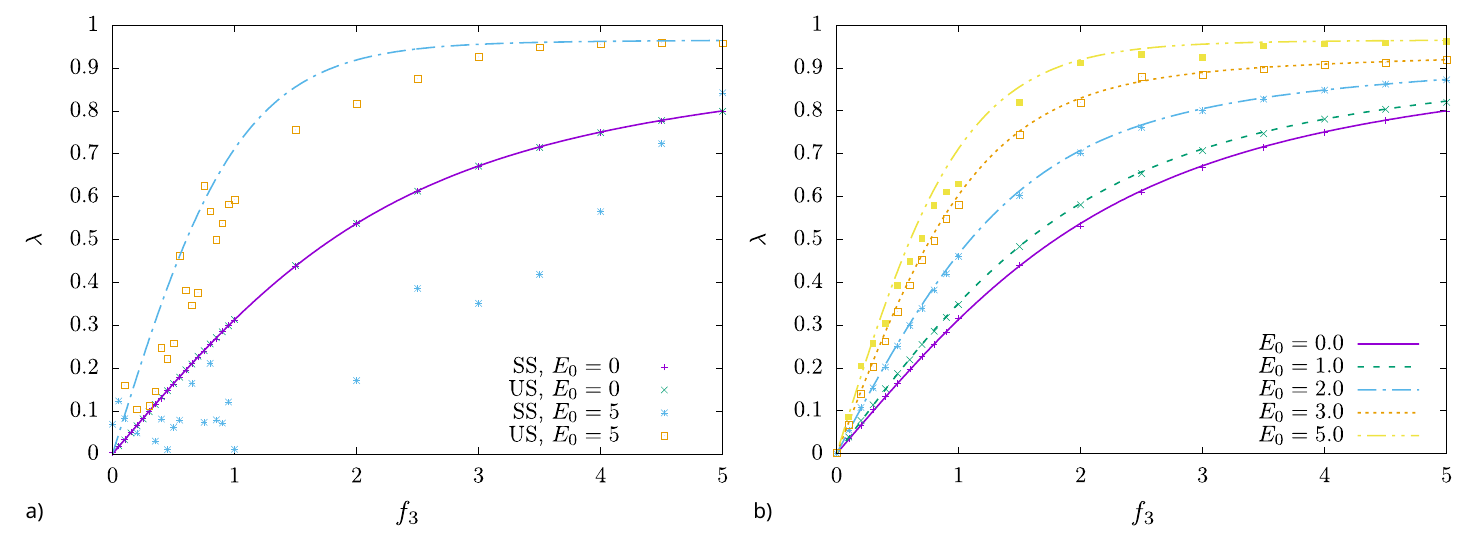}
	\caption{
		\captitle{Electroelasticity of a dielectric polymer chain with non-interacting monomers.}
		Force-stretch curves for dielectric polymer chains with $\invT = 1$ and $\susPara = 1$, and non-interacting monomers.
		a) compares the exact solution to MCMC simulations using standard (`SS') and umbrella (`US') sampling.
		While the predictions are nearly exact for $\emag = 0$, the agreement is poor for $\emag = 5$.
		This is due to MCMC localization for SS, and biasing the Markov chain toward ``unimportant'' samples for US.
		b) compares the exact solution to MCMC simulations using the symmetry-adapted clustering approach for $\emag = 0, 1, 2, 3,$ and $5$.
		The predictions and corresponding exact solutions are shown to agree well.
	}
	\label{fig:ni-comparison}
\end{figure*}

\Fref{fig:ni-comparison} compares the predicted stretch response to the known solution when interactions are negligible and the force is applied in the electric field direction.
\Fref{fig:ni-comparison}.a shows the results of MCMC simulations using standard (`SS') and umbrella (`US') sampling, where (following~\cite{grasingerIPforce}) the biasing potential for the umbrella sampling is chosen to be
\begin{equation}
  \weight{\mStateP; \fvec, \efield} = \begin{cases}
    \exp\left(\eta_1 \left(\eta_2 + \left(1 - \eta_2\right)e^{-\eta_3 \invT \fmag \mlen}\right)\sum_{i=1}^N \invT \um\left(\nvec_i\right)\right) & \susPara > \susPerp \\
    1 & \text{otherwise}
  \end{cases},
\end{equation}
where $\eta_1 \in \left[0, 1\right]$, $\eta_2 \in \left[0, 1\right]$, and $\eta_3 \geq 0$.
After various trials, values of $\eta_1 = 1.0$, $\eta_2 = 0.25$, and $\eta_3 = 1.0$ were found to be effective for small to moderate electric fields (i.e., $\emag \sqrt{\invT \left|\susPara - \susPerp\right|} \leq 3$).
While the predictions are nearly exact for $\emag = 0$, the agreement is poor for $\emag = 5$ and there is significant noise in the predictions for $\fmag \leq 1$.
For SS, this is because monomers become localized at $\nvec = \pm \edir$ and cannot traverse the energy barrier at $\nvec \cdot \edir = 0$.
For US, the chain traverses barriers but wastes steps sampling low-probability regions that contribute negligibly to ensemble averages. Both approaches struggle with the multimodal landscape: SS becomes trapped in energy wells, while US trades barrier-crossing ability for sampling efficiency. 
These issues persist even with $25$ replicas.
The symmetry-adapted clustering approach circumvents this trade-off. \Fref{fig:ni-comparison}b shows predictions using this method agree well with exact solutions for $\emag = 0, 1, 2, 3, 5$.

\subsection{Convergence rates}
We next examine how monomer-monomer interactions affect convergence of the various MCMC methods.
Since no analytical solution exists for the interacting case, we assess convergence by comparing running averages to final values.
For each sampling method, we again run $25$ replicas of $10^6$ steps.
The convergence rate for an observable, $\convRate{\obs}$, is defined as the slope obtained by fitting a line (via least squares) to $\log \sampleAvg{\left|\avg{\obs}_K - \sampleAvg{\avg{\obs}_{\nStates}}\right|}$ vs. $\log K$ data where $\sampleAvg{\square}$ denotes the sample average over replicas of $\square$ and $K$ is the step number.
The Central Limit Theorem for MCMC predicts $\convRate{\obs} = -1/2$ for well-mixing chains~\cite{brooks2011handbook,geyer2011introduction}; values less negative indicate slower-than-typical convergence due to poor mixing or autocorrelation.

We also define an error measure $\err{\obs}$ as the average deviation of individual replica averages from the mean over all replicas:
\begin{equation}
	\err{\obs} \coloneqq \frac{1}{\obs^{*}}\left(\sampleAvg{\left|\avg{\obs}_{\nStates} - \sampleAvg{\avg{\obs}_{\nStates}}\right|}\right),
\end{equation}
where $\obs^*$ is a reference value of the observable that serves to make the measure dimensionless.
This quantifies the expected error if only a single replica were used, providing a measure of sampling uncertainty.

Regarding observables of interest: let
\begin{equation}
	\pvec = \sum_{i=1}^{\numMonomers} \dipoleVec_i
\end{equation}
denote the net dipole of the polymer chain.
Let the coordinate system be Euclidean and such that $\ethree = \edir$; then $\Box_1 = \Box \cdot \eone$ and $\Box_2 = \Box \cdot \etwo$ are the components of $\Box$ orthogonal to the field direction, and $\Box_3 = \Box \cdot \ethree$ is the component of $\Box$ in the direction of $\edir$.
Then
\begin{equation}
	\observables = \set{\rmag_1, \rmag_2, \rmag_3, \pmag_1, \pmag_2, \pmag_3, \U, \rmag_1^2, \rmag_2^2, \rmag_3^2, \pmag_1^2, \pmag_2^2, \pmag_3^2, \U^2},
\end{equation}
are the observables of interest, and
$\rmag^{*} = \mlen$, $\pmag^{*} = \vperm \susPara \emag$, and $\UStar = \susPara \emag^2$.

The convergence rate and accuracy for standard (SS), umbrella (US), and symmetry-adapted (SA) sampling are compared in~\fref{tab:de-results-1} and \ref{tab:de-results-2} for various field strengths, forces, and temperatures. 
The symmetry-adapted clustering algorithm is the only approach to consistently perform well in all of the cases considered, both in terms of mitigating worst case observable convergence rate (i.e., $\max_{\obs \in \observables} \convRate{\obs}$) and error (i.e., $\max_{\obs \in \observables} \err{\obs}$).
In contrast, the non-symmetry-based approaches are only viable when the applied force has appropriate direction and magnitude to break symmetry sufficiently, and when $\invT$ is low enough (i.e., temperature is high enough).
For example, at $\invT = 1$ all methods perform reasonably, though even in this regime US shows large errors for some observables.
At high $\invT$ ($=10$) and/or high $\emag$ ($=5$) (\fref{tab:de-results-1} rows 2 and 3, respectively), maximum errors for SS and US increase significantly ($\sim 30 \times$ and $\sim 1900 \times$, respectively), while SA degrades only modestly ($\sim 6 \times$).
In some cases, US even exhibits positive $\convRate{\obs}$, indicating \emph{divergence}.
\begin{table*}
		\centering
		\setlength{\tabcolsep}{0.4em}
		\begin{tabular}{r|l l l l}
			\multicolumn{5}{c}{$\emag = 1, \: \fmag_1 = 0, \: \fmag_3 = 0, \: \invT = 1.$} \\
			\hline
			method & $\min_{\obs \in \observables} \convRate{\obs}$ & $\max_{\obs \in \observables} \convRate{\obs}$ & $\min_{\obs \in \observables} \err{\obs}$ & $\max_{\obs \in \observables} \err{\obs}$\\
			\hline
			SS & $-0.9940$ & $-0.3702$ & $+0.0298$ & $+0.7868$ \\
			US & $-0.2196$ & $-0.0884$ & $+0.9429$ & $+269.58$ \\
			SA  & $-0.5170$ & $-0.2227$ & $+0.0531$ & $+0.8753$ \\
			\multicolumn{5}{c}{ } \\
		\end{tabular}
		\begin{tabular}{r|l l l l}
			\multicolumn{5}{c}{$\emag = 1, \: \fmag_1 = 0, \: \fmag_3 = 0, \: \invT = 10.$} \\
			\hline
			method & $\min_{\obs \in \observables} \convRate{\obs}$ & $\max_{\obs \in \observables} \convRate{\obs}$ & $\min_{\obs \in \observables} \err{\obs}$ & $\max_{\obs \in \observables} \err{\obs}$\\
			\hline
			SS & $-0.8177$ & $+0.1826$ & $+0.0690$ & $+30.170$ \\
			US & $-0.0017$ & $+0.0006$ & $+2.2442$ & $+515578$ \\
			SA  & $-0.8383$ & $-0.2059$ & $+0.0879$ & $+5.2789$ \\
			\multicolumn{5}{c}{ } \\
		\end{tabular}
		\begin{tabular}{r|l l l l}
			\multicolumn{5}{c}{$\emag = 5, \: \fmag_1 = 0, \: \fmag_3 = 0, \: \invT = 1.$} \\
			\hline
			method & $\min_{\obs \in \observables} \convRate{\obs}$ & $\max_{\obs \in \observables} \convRate{\obs}$ & $\min_{\obs \in \observables} \err{\obs}$ & $\max_{\obs \in \observables} \err{\obs}$\\
			\hline
			SS & $-0.6788$ & $+0.0723$ & $+0.0546$ & $+21.403$ \\
			US & $-0.0000$ & $+0.0000$ & $+1.6537$ & $+1101.6$ \\
			SA  & $-1.2398$ & $-0.4721$ & $+0.0179$ & $+1.4278$ \\
			\multicolumn{5}{c}{ } \\
		\end{tabular}
    \caption{
    \captitle{MCMC performance under pure electric fields (no applied force).}
    Convergence rates and errors for standard (SS), umbrella (US), and symmetry-adapted (SA) sampling without applied forces.
    Columns show maximum/minimum convergence rates (more or less negative $\convRate{\obs}$) and errors ($\err{\obs}$) across observables.
    At low temperature ($\invT = 10$, row 2), energy barriers cause catastrophic failure of SS and US: maximum errors increase by factors of $38\times$ (SS) and $1900\times$ (US) relative to $\invT = 1$ (row 1), while SA degrades only $6\times$.
    High field strength ($\emag = 5$, row 3) similarly amplifies convergence difficulties for non-SA methods.
    }
	\label{tab:de-results-1}
\end{table*}

Applying force along the field direction ($f_3 = 1$, \fref{tab:de-results-2} row 1) reduces energy barriers, allowing SS and US to converge. However, force perpendicular to the field ($f_1 = 1$, \fref{tab:de-results-2} row 2) does not help US (max error $\sim 1700$), suggesting the effectiveness of symmetry breaking depends on force orientation. 
When large forces are applied in both directions ($f_1 = f_3 = 3$, row 3), the discrete symmetry is sufficiently broken that all methods perform comparably.

Across conditions, US performance varies dramatically: either comparable to SA (max error $\sim 0.15$) or catastrophically poor (max error $> 10^5$). 
This reflects US's sensitivity to the choice of biasing potential -- which is difficult to construct robustly in general and may not exist at all for certain energy landscapes. When discrete symmetries are well-understood, a symmetry-adapted clustering approach may provide a more reliable strategy for overcoming energy barriers without manual tuning of bias functions.
\begin{table*}
		\centering
		\setlength{\tabcolsep}{0.4em}
		\begin{tabular}{r|l l l l}
			\multicolumn{5}{c}{$\emag = 1, \: \fmag_1 = 0, \: \fmag_3 = 1, \: \invT = 10.$} \\
			\hline
			method & $\min_{\obs \in \observables} \convRate{\obs}$ & $\max_{\obs \in \observables} \convRate{\obs}$ & $\min_{\obs \in \observables} \err{\obs}$ & $\max_{\obs \in \observables} \err{\obs}$\\
			\hline
			SS & $-1.4812$ & $-0.5055$ & $+0.0163$ & $+1.5976$ \\
			US & $-0.5297$ & $-0.1147$ & $+0.0727$ & $+0.1510$ \\
			SA  & $-1.3466$ & $-0.5288$ & $+0.0146$ & $+0.2882$ \\
			\multicolumn{5}{c}{ } \\
		\end{tabular}
		\begin{tabular}{r|l l l l}
			\multicolumn{5}{c}{$\emag = 1, \: \fmag_1 = 1, \: \fmag_3 = 0, \: \invT = 10.$} \\
			\hline
			method & $\min_{\obs \in \observables} \convRate{\obs}$ & $\max_{\obs \in \observables} \convRate{\obs}$ & $\min_{\obs \in \observables} \err{\obs}$ & $\max_{\obs \in \observables} \err{\obs}$\\
			\hline
			SS & $-0.9514$ & $-0.3526$ & $+0.1285$ & $ +0.8324$ \\
			US & $-0.0000$ & $+0.0000$ & $+1.9725$ & $+1717.2384$ \\
			SA  & $-1.0133$ & $-0.4446$ & $+0.1042$ & $+0.9607$ \\
			\multicolumn{5}{c}{ } \\
		\end{tabular}
		\begin{tabular}{r|l l l l}
			\multicolumn{5}{c}{$\emag = 3, \: \fmag_1 = 3, \: \fmag_3 = 3, \: \invT = 1.$} \\
			\hline
			method & $\min_{\obs \in \observables} \convRate{\obs}$ & $\max_{\obs \in \observables} \convRate{\obs}$ & $\min_{\obs \in \observables} \err{\obs}$ & $\max_{\obs \in \observables} \err{\obs}$\\
			\hline
			SS & $-1.4082$ & $-0.7502$ & $+0.0398$ & $+0.1001$ \\
			US & $-0.7065$ & $-0.4341$ & $+0.0580$ & $+0.1551$ \\
			SA  & $-1.1827$ & $-0.4405$ & $+0.0324$ & $+0.1500$ \\
			\multicolumn{5}{c}{ } \\
		\end{tabular}
	\caption{
    \captitle{MCMC performance with combined fields and forces.}
    Convergence rates and errors for standard (SS), umbrella (US), and symmetry-adapted (SA) sampling with applied field and forces.
    Force parallel to field ($f_3$, row 1) enables SS/US convergence; perpendicular force ($f_1$, row 2) does not rescue US.
    Large forces in both directions (row 3) break symmetry sufficiently that all methods perform comparably.
    }
	\label{tab:de-results-2}
\end{table*}

\subsection{Electro-elasticity}

\begin{figure*}
	\centering
	\includegraphics[width=\linewidth]{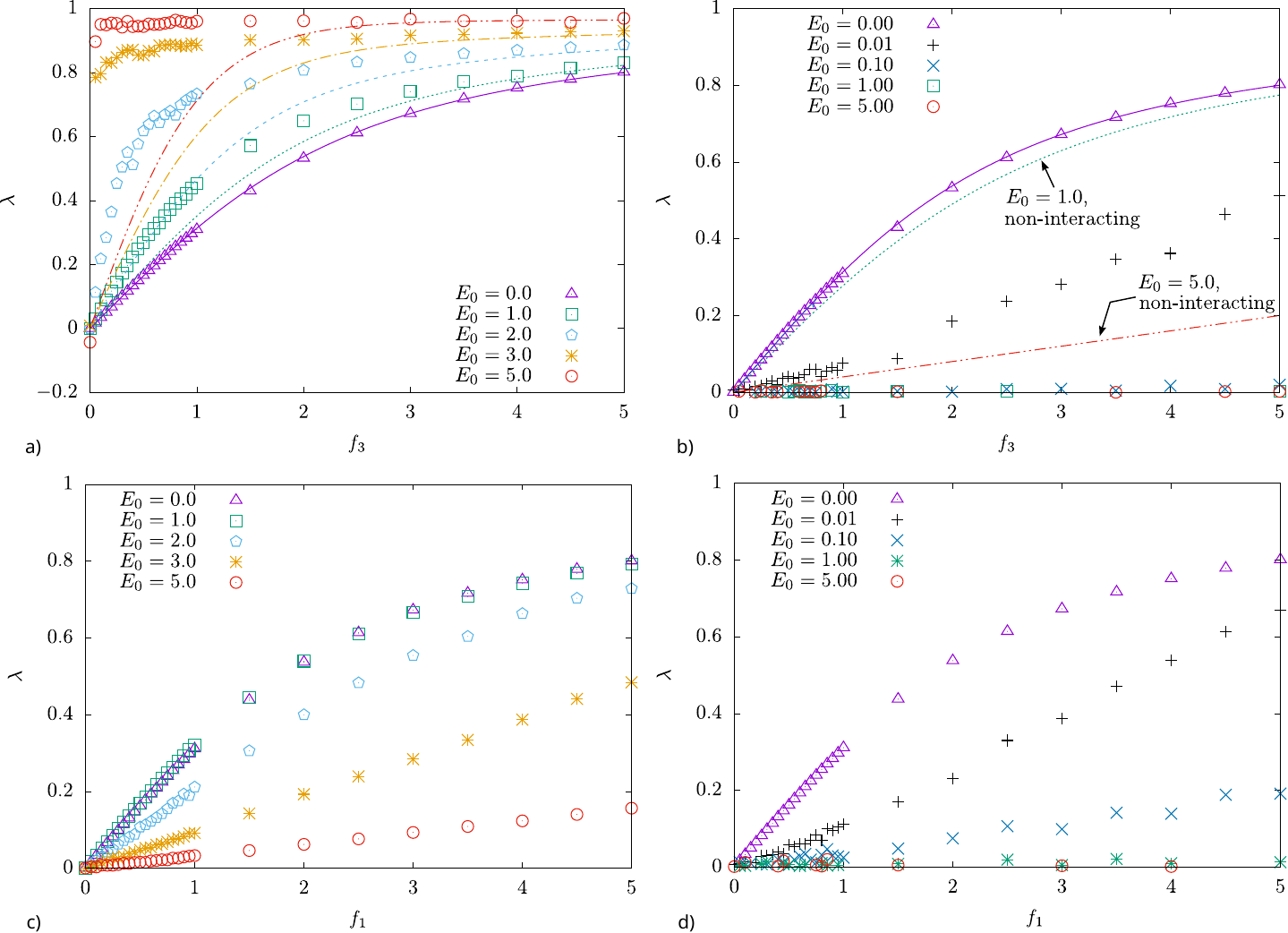}
	\caption{
		\captitle{Electroelasticity of a dielectric polymer chain with monomer-monomer interactions.}
		Top row: the force is applied parallel to the field for (a) field-aligning (FA) monomers ($\susPara = 1, \susPerp = 0$) and (b) field-disaligning (FD) monomers ($\susPara = 0, \susPerp = 1$). 
        Lines represent the analytical solutions for non-interacting chains, while markers denote simulated chains with dipole-dipole interactions. 
        Bottom row: the force is applied orthogonal to the field for (c) FA chains and (d) FD chains. 
        The inclusion of dipole-dipole interactions leads to electrically-driven tautness in FA chains at moderate to high fields (panel a) and a highly sensitive, orthogonal chain collapse in FD chains at low fields (panels b \& d).
	}
	\label{fig:FxFz}
\end{figure*}

Having established the symmetry-adapted sampling for dielectric polymer chains, we now examine the influence of dipole-dipole interactions on chain electroelasticity.
\Fref{fig:FxFz} shows force-extension curves for field-aligning (FA) and field-disaligning (FD) chains under parallel and perpendicular loading configurations.
For simplicity, the dipole susceptibility of FA chains and FD chains were taken to be $\susPara = 1, \susPerp = 0$ and $\susPara = 0, \susPerp = 1$, respectively.
The parameters $\invT = 1$, $\vperm = 1$, and $\numMonomers = 100$ were all constant across simulations.
A burn-in schedule of $\invT = \left\{10^{-3}, 10^{-2}, 10^{-1}, 0.5, 1\right\}$ was used where $10^5$ steps were run for each temperature level. 
Each set of system parameters was run with $5$ replicas and $2.5 \times 10^6$ steps after burn-in (total: $12.5 \times 10^6$ samples per parameter set).

\paragraph*{Force-stretch along the electric field direction.}
The top row illustrates the electroelastic response when the force is applied along the field direction. For the FA chains shown in panel a), an analytical solution exists for the non-interacting case (indicated by lines), which isolates the effect of dipole-dipole interactions (indicated by markers). 
With interactions, the FA chain stretches significantly more under the same electric field and force compared to its non-interacting counterpart. 
This disparity grows as the electric field increases. 
At large field strengths, such as $\emag=5.0$, the chain exhibits an electrically-driven tautness, stretching fully in the field direction even as the applied force vanishes, displaying almost no tension stiffness (and possibly a tangent compressive stiffness, since the chain must pass through an energy barrier to snap in the opposite direction of the field).
This behavior is primarily driven by dipole-dipole interactions that cause neighboring monomers to align with one another. 
However, as the mechanical force becomes sufficiently large, mechanical energy dominates over electrostatic energy; all curves eventually converge toward the high-force mechanical limit, though the highest fields plotted require significantly larger forces to reach this convergence.

Conversely, for the field-disaligning (FD) chains shown in panel b), dipole-dipole interactions cause the chains to stretch less than the non-interacting baseline. 
This chain stiffening stems from two monomer-scale phenomena: first, dipoles tend to form orthogonal to the monomer axis, driving individual monomers to orient orthogonally to the applied field; second, neighboring monomers favorably ``fold in'' relative to each other. 
Because the dipoles form orthogonal to the monomer axis, this folding aligns the dipoles nearly end-to-end, creating a highly energetically favorable state. 
Together, these effects cause the chain to effectively ``collapse'' in the plane orthogonal to the field, a phenomenon consistent with previous statistical field theoretic simulations~\cite{khandagale2024statistical,khandagale2025nonlocal}. 
Notably, there is an asymmetry in sensitivity between the two chain types. 
This electrically-driven chain collapse alters the chain's electroelasticity at very low fields, around $\emag=0.01$ to $\emag=0.10$. 
In contrast, the FA tautness requires fields roughly one to two orders of magnitude higher (e.g., $\emag=1.0$ to $\emag=5.0$ ) to become pronounced, highlighting how much more energetically favorable the orthogonal folding mechanism is compared to parallel alignment.

\paragraph*{Force-stretch orthogonal to field direction.}
The bottom row depicts the chain response when forced orthogonal to the electric field direction. 
For the FA chains in panel c), the effective stiffness increases with the applied field. 
In this configuration, a competition emerges: the monomers naturally want to align with the electric field along the orthogonal axis, while the force acts to pull them perpendicular to it. 
Consequently, no structural collapse occurs. 
In contrast, for the FD chains in panel d), a phenomenon similar to the aforementioned collapse is observed, though it is markedly less pronounced. 
For instance, there is still finite stretch at the applied forces for intermediate fields like $\emag=0.10$, indicating that the force partially disrupts the end-to-end dipole folding that drives the complete collapse seen when forced in the parallel direction.

\subsection{Dielectric response}
\begin{figure}
	\centering
	\includegraphics[width=0.65\linewidth]{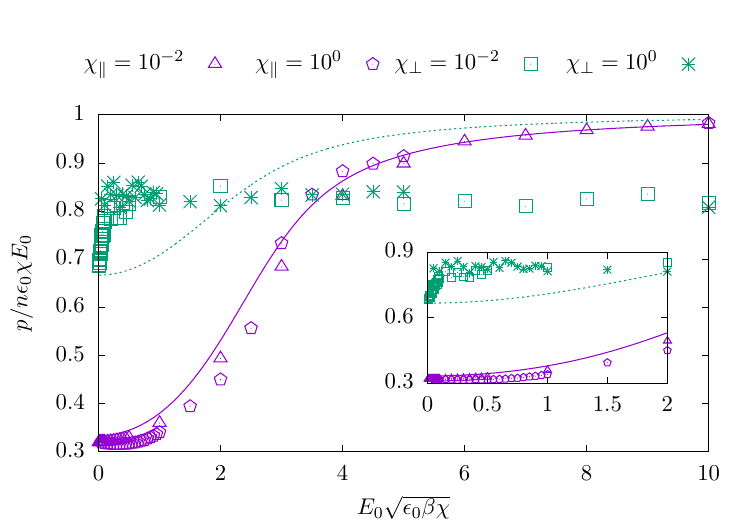}
	\caption{
		\captitle{Dielectric response of a polymer chain with monomer-monomer interactions.}
        Normalized dielectric response of field-aligning (FA, purple) and field-disaligning (FD, green) chains as a function of dimensionless applied electric field. Solid (FA) and dashed (FD) lines denote the analytical approximation in which dipole-dipole interactions are neglected. Symbols denote MCMC results at two dipole susceptibility magnitudes: $10^{-2}$ ($\triangle$, FA; $\square$, FD) and $10^{0}$ (\pentagon, FA; $*$, FD). Dipole-dipole interactions have a negligible effect on the FA dielectric response across both susceptibilities, whereas FD chains exhibit a sharper initial rise and an asymptotic saturation well below unity.
	}
	\label{fig:muz}
\end{figure}

\Fref{fig:muz} presents the normalized dielectric response, $\pmag / \numMonomers \vperm \chi_{\Box} \emag$, for FA and FD chains as a function of the dimensionless applied electric field $\tilde{\emag} = \emag\sqrt{\vperm \invT \chi_{\Box}}$. The chain polarization is normalized by its maximum value such that unity corresponds to full saturation. The dimensionless field strength captures the competition between electrostatic alignment of dipoles with the applied field and thermal randomization, ensuring that variations in susceptibility isolate the effect of dipole-dipole interactions from all other energetics. Two susceptibility magnitudes are considered for each chain type: $10^{-2}$ and $10^{0}$, spanning two orders of magnitude in the strength of dipole-dipole coupling. The solid lines denote the corresponding analytical approximation in which dipole-dipole interactions are neglected entirely.

For the FA chains, dipole-dipole interactions have a negligible effect on the dielectric response: the analytical curve, the low-susceptibility MCMC results, and the high-susceptibility MCMC results all roughly agree. A slight suppression of the dielectric response is visible in certain regimes, particularly at small applied fields. This can be understood as follows: in the absence of interactions, configurations containing a mixture of monomers aligned with and opposed to the field direction carry low electrostatic energy and high entropy, and therefore contribute appreciably to the polarization. When interactions are included, however, the FA chain penalizes such configurations: a monomer aligned with the field adjacent to one folded against it produces two dipoles that are near in space, side by side, and aligned. This has a high interaction energy, rendering these mixed configurations less probable and marginally suppressing the net polarization.

The situation for FD chains is qualitatively different. Without dipole-dipole interactions, the normalized polarization begins at $2/3$ and gradually saturates to unity at large fields. With interactions, the initial increase from $2/3$ is considerably sharper, but the polarization asymptotes to $\approx0.8$; full saturation is never achieved. This is likely due to the geometric constraint imposed by saturation: for the FD polarization to reach unity, all monomers must orient within the plane orthogonal to the applied field, which again places neighboring dipoles side by side and aligned. The resulting interaction energy is large, making these fully saturated configurations energetically unfavorable. The net effect is that dipole-dipole interactions are far more consequential for the FD dielectric response than for the FA.

This distinction is notable because some of the broader trends from electroelasticity do not carry over to the dielectric response. For instance, dipole-dipole interactions are negligible for the FA dielectric response but are significant for the FA force-stretch relationship, where they lead to chain tautening. The dielectric and mechanical responses thus are modulated by different aspects of the interplay between applied field and dipole-dipole coupling, and conclusions drawn from one cannot be straightforwardly extrapolated to the other.

\subsection{Monomer-monomer alignment and order}
The Hermans orientation parameter is a metric widely utilized as an order parameter for systems of constituents that have an axial symmetry (e.g., liquid crystal community).
Here it takes the form, \begin{equation}
	S = \frac{3 \langle \langle \cos^2 \theta_i \rangle \rangle_i - 1}{2},
\end{equation}
where $\langle \Box \rangle_i$ denotes an average of $\Box$ over monomers $i = 1, \dots, \numMonomers$.
This parameter characterizes the average alignment of the monomers relative to a specific reference axis, with values ranging from $-0.5$ to $1.0$. A value of $1.0$ represents perfect parallel alignment along the axis, while $-0.5$ indicates that the orientations are saturated within the plane normal to the axis; an isotropic distribution of orientations corresponds to $0$. 

\begin{figure*}
	\centering
	\includegraphics[width=\linewidth]{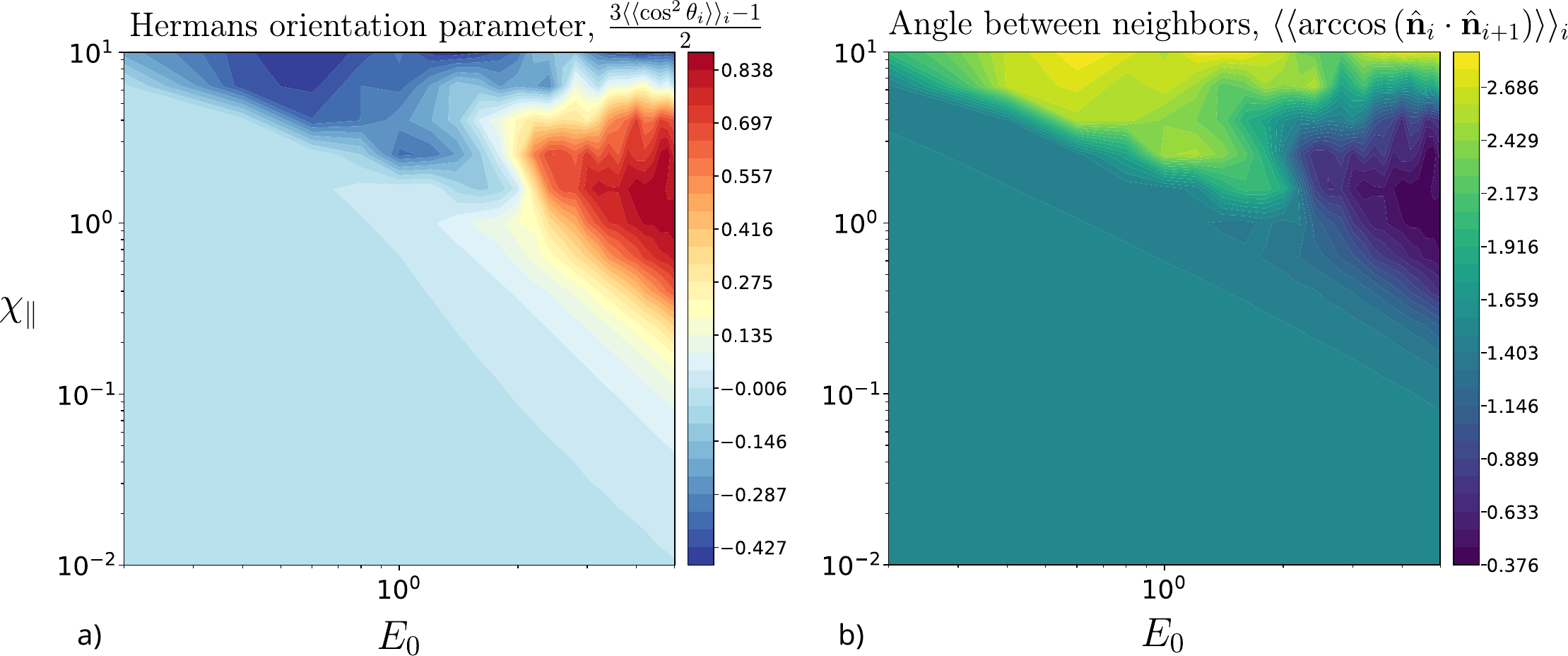}
	\caption{
		a) Hermans orientation parameter and b) mean angle between neighboring monomers for FA chains as a function of $\emag$ and $\susPara$.
        At low $\emag$ and $\susPara$, the orientation parameter is near $0$ and the neighbor angle is near $\pi / 2$, consistent with an isotropic monomer distribution.
        Two regimes emerge at higher susceptibility: 1) the orientation parameter approaches $-1/2$ (transverse-plane confinement) at low to moderate $\emag$, accompanied by neighbor anti-alignment ($\psi > \pi / 2$), and 2) transitions sharply toward $1$ (field-axis alignment) at high $\emag$, with neighbors becoming co-aligned ($\psi < \pi / 2$).
	}
	\label{fig:K1_orientation}
\end{figure*}

\paragraph*{Field-aligning chains.}
\Fref{fig:K1_orientation} shows a) the Hermans orientation parameter, and b) the mean angle between neighboring monomers, $\psi \coloneq \langle \langle \arccos\left(\nvec_i \cdot \nvec_{i+1}\right) \rangle \rangle_i$, where both are given for FA chains across the $\left(\emag, \susPara\right)$ parameter space. The parameters $\invT = 1$, $\vperm = 1$, $\fvec = \nullvec$, and $\numMonomers = 100$ were all constant across simulations. (A burn-in schedule of $\invT = \left\{10^{-3}, 10^{-2}, 10^{-1}, 0.5, 1\right\}$ was used where $10^5$ steps were run for each temperature level.  Each set of system parameters was run with $5$ replicas and $2.5 \times 10^6$ steps after burn-in (total: $12.5 \times 10^6$ samples per parameter set)). At low $\emag$ and low $\susPara$, the Hermans orientation parameter is nearly $0$ (\Fref{fig:K1_orientation}a), indicating an approximately isotropic distribution of monomer orientations. This is expected: in this regime both dipole-field and dipole-dipole interactions are weak relative to thermal energy, and, when also in the absence of applied forces, there is no driver for orientational preference.

At moderate to high susceptibility, two distinct regimes emerge. When the susceptibility is large but $\emag$ is low to moderate, the Hermans orientation parameter approaches $-1/2$, the theoretical minimum, indicating that monomer orientations are saturating in the plane orthogonal to the applied field. This can be understood through the structure of the FA induced dipole, $\dipoleVec_i = \susPara \left( \efield \cdot \nvec_i \right) \nvec_i$: the dipole magnitude is proportional to the projection of the monomer axis onto the field direction, so rotating into the transverse plane reduces the induced dipole and thereby the electrostatic energy. At large $\emag$, the orientation parameter transitions sharply toward unity, indicating alignment along the field axis. Here the dipole-field interaction energy, which scales with $\emag^2 \susPara$, is sufficient to overcome the energetic penalty associated with the large co-aligned dipoles that result from axial orientation.

The mean neighbor angle (\Fref{fig:K1_orientation}b) reveals complementary microstructural detail. In the high-susceptibility, low-to-moderate-$\emag$ regime where $S \rightarrow -1/2$, the neighbor angle significantly exceeds $\pi / 2$, the value expected for uncorrelated orientations. This means that neighboring monomers are actively anti-aligning (i.e., the chain is folding locally) within the transverse plane. Although monomers in this regime are predominantly orthogonal to the field, thermal fluctuations produce small out-of-plane tilts and correspondingly small residual dipoles. Anti-aligned neighbors orient these residual dipoles in opposing directions, which is energetically favorable. The anti-alignment of neighbors is thus driven by dipole-dipole interactions even though the dipole magnitudes themselves are small.

When $\emag$ is large, the neighbor angle drops sharply, indicating that consecutive monomers become co-aligned and the chain locally extends along the field direction. This is consistent with the jump in the Hermans parameter toward unity. Notably, local folding -- alternating monomer orientations along the field axis -- provides no relief from high energy dipole-dipole interactions in this regime. Because $\dipoleVec_i = \susPara \left( \efield \cdot \nvec_i \right) \nvec_i$ is invariant under $\nvec_i \rightarrow -\nvec_i$, reversing the monomer direction does not change the induced dipole, and zigzagging along the field axis yields co-aligned, side-by-side dipoles with high interaction energy. The chain therefore has no incentive to fold locally and instead locally extends. Even a small applied force in this regime would result in a jump in end-to-end vector length such that the chain is taut.

\begin{figure*}
	\centering
	\includegraphics[width=\linewidth]{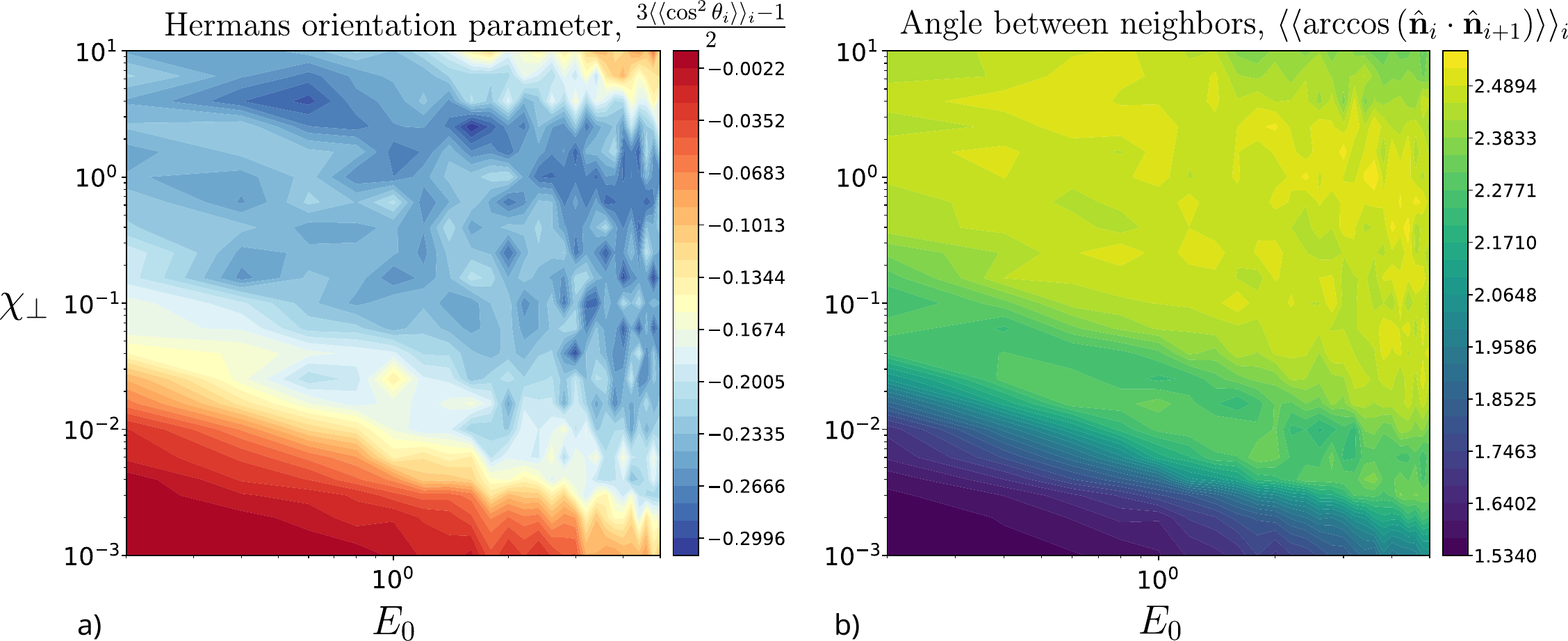}
	\caption{a) Hermans orientation parameter and b) mean angle between neighboring monomers for FD chains as a function of $\emag$ and $\susPerp$.
    At low $\emag$ and $\susPerp$, the monomer distribution is approximately isotropic.
    At moderate $\emag \susPerp$, the orientation parameter transitions sharply to the $-0.3$ to $-0.2$ range, and the neighbor angle increases toward $\approx 0.8 \pi$, indicating transverse-plane orientation and local chain folding, though neither quantity reaches its theoretical limit.
    Visible noise in the orientation parameter in this regime may reflect a rugged energy landscape or flattened free energy.
    At high $\emag$ and $\susPerp$, the orientation parameter trends back toward $0$, possibly suggesting a complex competition between dipole-field and dipole-dipole interactions.
	}
	\label{fig:K2_orientation}
\end{figure*}

\paragraph*{Field-disaligning chains.}
To disentangle the competing mechanisms underlying the distinct electroelastic and dielectric responses, we turn to characterization of monomer orientational order.
\Fref{fig:K2_orientation} presents the corresponding orientation diagnostics for FD chains. At low $\emag$ and low $\susPerp$, the Hermans orientation parameter is approximately $0$ (\Fref{fig:K2_orientation}a) and the mean neighbor angle is near $\pi/2$ (\Fref{fig:K2_orientation}b), consistent with an isotropic distribution of monomer orientations in a regime where both dipole-field and dipole-dipole interactions are weak relative to thermal energy.

When the product $\emag \susPerp$ exceeds roughly $10^{-2}$, the orientation parameter transitions sharply to the $-0.3$ to $-0.2$ range, indicating a significant preference for monomer orientations in the plane orthogonal to the applied field. However, the orientation parameter does not approach the saturation value of $-1/2$, in contrast to the FA chains where near-saturation was observed at high $\susPerp$. The orientation parameter also exhibits visible noise in this regime, which may reflect either a rugged energy landscape with many local minima and correspondingly slow statistical convergence, or a flattening of the free energy with respect to the orientation parameter such that thermodynamic fluctuations of the orientation parameter become significant. These two scenarios are not mutually exclusive and may be difficult to distinguish without comprehensive analysis of the MCMC convergence diagnostics.

Concurrently, the mean neighbor angle in the moderate-to-high $\emag$-$\susPerp$ regime increases well beyond $\pi/2$, reaching $\approx 0.8\pi$ (\Fref{fig:K2_orientation}b). This indicates that the chain is folding locally, with neighboring monomers adopting obtuse mutual angles. The neighbor angle does not, however, reach $\pi$, which would correspond to complete reversal of consecutive monomer orientations. This partial folding likely reflects the same competition between dipole-field and dipole-dipole interactions that prevents saturation of the orientation parameter: configurations that minimize one interaction term do not generally minimize the other.

At high $\emag$ and high $\susPerp$, the orientation parameter trends back toward $0$. We speculate that this reflects a complex competition between the dipole-field and dipole-dipole interactions. The dipole-field interaction favors confinement of all monomer orientations to the plane orthogonal to the applied field, which would drive $S \rightarrow -1/2$. However, in that configuration the FD induced dipoles $\dipoleVec_i = \susPerp\left(\efield - \left(\efield \cdot \nvec_i\right)\nvec_i\right)$ are all aligned with the applied field and spatially adjacent, producing large interaction energies. The system is thus unable to fully exploit the transverse plane, and the orientation parameter is pushed back toward less negative values. This mechanism is consistent with the incomplete saturation of the FD dielectric response observed in \Fref{fig:muz}.

\section{Conclusion} \label{sec:conclusion}
\paragraph*{Summary.} This work developed a discrete-symmetry-adapted Markov chain Monte Carlo method for sampling the statistical mechanics of dielectric polymer chains subject to combined mechanical forces and electric fields. The method exploits the fact that the dipole-dipole interaction energy between neighboring monomers is invariant under simultaneous reflection of their orientations about the plane orthogonal to the applied field. By constructing variable-sized clusters of neighboring monomers and reflecting them collectively, the algorithm achieves rapid mixing between symmetric energy wells without biasing potentials or temperature replicas. The symmetry-adapted approach consistently outperforms standard and umbrella sampling across all parameter regimes considered, and renders feasible simulations that are otherwise intractable due to critical slowing down.
Application of the method to dielectric polymer chains with monomer-monomer interactions revealed qualitatively distinct electroelastic and orientational phenomena depending on the anisotropy of the monomer polarizability. For field-aligning chains, dipole-dipole interactions drive local chain straightening, an apparent compressive stiffness, and electrically induced tautness at moderate to high fields. For field-disaligning chains, dipole-dipole interactions instead promote local chain folding and an electrically induced collapse. These phenomena were quantified through the Hermans orientation parameter and the mean angle between neighboring monomers, mapped across the $\left(\emag, \chi_\Box\right)$ parameter space. The resulting phase diagrams (\Fref{fig:K1_orientation} and \Fref{fig:K2_orientation}) exhibit sharp transitions in orientational order that are suggestive of underlying phase transitions. For the FD chains, the orientational order transitions correspond directly with a sharp jump in the chain polarization, linking microstructural rearrangement to the macroscopic dielectric response.

\paragraph*{Limitations.} The present work is subject to several limitations. Dipole-dipole interactions were restricted to nearest neighbors along the chain backbone for computational tractability, neglecting longer-range electrostatic coupling between monomers that are spatially proximate due to chain folding but topologically distant. Inclusion of these long-range interactions may give rise to additional orientational order phenomena and corresponding electroelastic effects not captured here. Excluded volume effects were also neglected, which may become significant precisely in the collapsed-chain configurations where monomers are densely packed.

\paragraph*{Outlook.} These limitations present clear opportunities for future work. Designing symmetry-adapted clustering algorithms that remain efficient in the presence of long-range interactions is a nontrivial challenge, as the interaction energy is no longer invariant under reflection of a local cluster when non-neighboring contributions are included. A more comprehensive investigation of the apparent phase transitions, particularly with long-range electrostatics, would also be valuable for establishing whether the sharp transitions observed here sharpen into true thermodynamic singularities or remain crossovers. More broadly, the statistical mechanics formulations for many multifunctional elastomers share a similar structure: field-monomer and monomer-monomer interaction energies that depend on even powers of $\nvec$, combined with an applied force term that is linear in $\nvec$ (e.g., liquid crystal elastomers~\cite{de1993physics,warner2007liquid}). In such systems, clustering-type symmetry-adapted MCMC with inversion ($\nvec \rightarrow -\nvec$) or reflection actions about field-induced symmetry planes may offer a similarly promising route to efficient sampling.

The interplay between dipole-field and dipole-dipole interactions in dielectric polymer chains produces a rich landscape of electroelastic and orientational phenomena -- from tautness to collapse, and gradual to sharp jumps in polarization -- that cannot be anticipated from non-interacting models alone. Symmetry-adapted sampling methods provide both the computational efficiency and the physical insight needed to navigate this landscape.

\section*{Software availability}
The code(s) used for analysis and generation of data for this work is available at \url{https://github.com/grasingerm/polymer-stats}.

\section*{Acknowledgments}
The author acknowledges the support of the Air Force Research Laboratory.

\bibliographystyle{unsrtnat}
\bibliography{master}

\end{document}

%% file: preamble.tex
\usepackage[utf8]{inputenc}
\usepackage{mathtools}
\usepackage{amsmath}
\usepackage{amsthm}
\usepackage{amsbsy}
\usepackage{amssymb}
\usepackage{amscd}
\usepackage{amsfonts}
\usepackage{bm}
\usepackage{graphics}
\usepackage{graphicx}
\usepackage[plain]{fancyref}
\usepackage{soul}
\usepackage{hyperref}
\setcitestyle{numbers,sort&compress}
\usepackage{verbatim}
\usepackage[margin=25mm]{geometry}
\usepackage{paralist}
\usepackage{wasysym}
\usepackage{MnSymbol}
\graphicspath{ {./figs/} }

\newcommand*{\fancyrefapplabelprefix}{app}
\frefformat{plain}{\fancyrefapplabelprefix}{appendix~#1}
\Frefformat{plain}{\fancyrefapplabelprefix}{Appendix~#1}
\frefformat{main}{\fancyrefapplabelprefix}{appendix~#1}
\Frefformat{main}{\fancyrefapplabelprefix}{Appendix~#1}

\include{notation}

\makeatletter
\newcommand*{\gnuplotinput}[2][1.0]{%
  \begingroup
  \let\@gnplt@input@includegraphics=\includegraphics
  \def\includegraphics##1{\@gnplt@input@includegraphics[scale=#1]{#2}}%
  \let\@gnplt@input@picture=\picture
  \def\picture{\unitlength=#1\unitlength\relax\@gnplt@input@picture}%
  \input{#2}%
  \endgroup
}
\makeatother

\theoremstyle{remark}

\newcommand*{\fancyrefproplabelprefix}{prop}
\frefformat{plain}{\fancyrefproplabelprefix}{proposition~#1}
\Frefformat{plain}{\fancyrefproplabelprefix}{Proposition~#1}
\frefformat{main}{\fancyrefproplabelprefix}{proposition~#1}
\Frefformat{main}{\fancyrefproplabelprefix}{Proposition~#1}

\newcommand*{\fancyrefremlabelprefix}{rem}
\frefformat{plain}{\fancyrefremlabelprefix}{remark~#1}
\Frefformat{plain}{\fancyrefremlabelprefix}{Remark~#1}
\frefformat{main}{\fancyrefremlabelprefix}{remark~#1}
\Frefformat{main}{\fancyrefremlabelprefix}{Remark~#1}

\newcommand*{\fancyrefdeflabelprefix}{def}
\frefformat{plain}{\fancyrefdeflabelprefix}{definition~#1}
\Frefformat{plain}{\fancyrefdeflabelprefix}{Definition~#1}
\frefformat{main}{\fancyrefdeflabelprefix}{definition~#1}
\Frefformat{main}{\fancyrefdeflabelprefix}{Definition~#1}

\newcommand*{\fancyreflemlabelprefix}{lem}
\frefformat{plain}{\fancyreflemlabelprefix}{lemma~#1}
\Frefformat{plain}{\fancyreflemlabelprefix}{Lemma~#1}
\frefformat{main}{\fancyreflemlabelprefix}{lemma~#1}
\Frefformat{main}{\fancyreflemlabelprefix}{Lemma~#1}

\newcommand*{\fancyrefasslabelprefix}{ass}
\frefformat{plain}{\fancyrefasslabelprefix}{assumption~#1}
\Frefformat{plain}{\fancyrefasslabelprefix}{Assumption~#1}
\frefformat{main}{\fancyrefasslabelprefix}{assumption~#1}
\Frefformat{main}{\fancyrefasslabelprefix}{Assumption~#1}

\newcommand*{\fancyrefconvlabelprefix}{conv}
\frefformat{plain}{\fancyrefconvlabelprefix}{convention~#1}
\Frefformat{plain}{\fancyrefconvlabelprefix}{Convention~#1}
\frefformat{main}{\fancyrefconvlabelprefix}{convention~#1}
\Frefformat{main}{\fancyrefconvlabelprefix}{Convention~#1}

\newcommand{\captitle}[1]{\emph{#1}}

\usepackage{color}
\definecolor{myorange}{rgb}{1, 0.647, 0}

%% file: notation.tex
\newcommand{\invT}{\beta}
\newcommand{\U}{U}
\DeclareMathOperator{\sign}{sgn}

\newcommand{\group}{\mathcal{G}}
\newcommand{\genBy}[1]{\left\langle\left\langle #1 \right\rangle\right\rangle}
\newcommand{\UGroup}{\U_{\group}}
\newcommand{\UO}{\U_{0}}
\newcommand{\otherGroup}{G}
\newcommand{\groupWeight}{\omega}

\newcommand{\reflAbout}[1]{\sigma_{#1}}
\newcommand{\iden}{e}

\newcommand{\probSym}{\pi}
\newcommand{\prob}[1]{\probSym\left(#1\right)}
\newcommand{\modProb}[1]{\probSym'\left(#1\right)}
\newcommand{\transProb}[2]{p\left(#1 \rightarrow #2\right)}
\newcommand{\accProb}[2]{a\left(#1 \rightarrow #2\right)}
\newcommand{\trialProb}[2]{t\left(#1 \rightarrow #2\right)}

\newcommand{\clusterProbSym}{\zeta}
\newcommand{\clusterProb}[1]{\clusterProbSym\left(#1\right)}
\newcommand{\pSpace}{\Gamma}
\newcommand{\mState}{x}
\newcommand{\mStar}{\mState^{*}}
\newcommand{\otherMState}{y}
\newcommand{\df}[1]{\mathrm{d}#1}
\newcommand{\weightSym}{w}
\newcommand{\weight}[1]{\weightSym\left(#1\right)}
\newcommand{\avg}[1]{\left\langle#1\right\rangle}

\newcommand{\obs}{O}
\newcommand{\nStates}{\mathcal{K}}
\newcommand{\chain}{\mathcal{C}}
\newcommand{\chainAvg}[1]{\avg{#1}_{\chain}}
\newcommand{\modChainAvg}[1]{\avg{#1}_{\chain'}}

\newcommand{\groupAct}[2]{#1 \cdot #2}

\newcommand{\hastings}[2]{\mathcal{H}\left(#1 \rightarrow #2\right)}

\newcommand{\fmag}{f}
\newcommand{\fvec}{\mathbf{\fmag}}
\newcommand{\xmag}{x}
\newcommand{\xvec}{\mathbf{\xmag}}

\newcommand{\set}[1]{\left\{ #1 \right\}}
\newcommand{\setSize}[1]{\left|#1\right|}

\newcommand{\nullvec}{\mathbf{0}}
\newcommand{\smallParam}{\epsilon}

\newcommand{\UStar}{\U^{*}}

\newcommand{\sampleAvg}[1]{\overline{#1}}
\newcommand{\convRate}[1]{\alpha_{#1}}
\newcommand{\observables}{\mathbb{O}}

\newcommand{\err}[1]{\mathcal{E}_{#1}}

\newcommand{\eone}{\hat{\mathbf{e}}_1}
\newcommand{\etwo}{\hat{\mathbf{e}}_2}
\newcommand{\ethree}{\hat{\mathbf{e}}_3}

\newcommand{\vperm}{\epsilon_0}
\newcommand{\nvec}{\hat{\mathbf{n}}}
\newcommand{\rmag}{r}
\newcommand{\rvec}{\mathbf{\rmag}}

\newcommand{\emag}{E_0}
\newcommand{\efield}{\mathbf{E}_0}
\newcommand{\edir}{\hat{\mathbf{E}}_0}
\newcommand{\dipoleMag}{\mu}
\newcommand{\dipoleVec}{\bm{\mu}}
\newcommand{\susSym}{\chi}

\newcommand{\susPara}{\susSym_{\parallel}}
\newcommand{\susPerp}{\susSym_{\perp}}
\newcommand{\posVec}[1]{\mathbf{q}_{#1}}
\newcommand{\relPosVec}[1]{\mathbf{q}_{#1}}
\newcommand{\relPosMag}[1]{q_{#1}}
\newcommand{\relDirVec}[1]{\hat{\mathbf{q}}_{#1}}
\newcommand{\mlen}{b}
\newcommand{\numMonomers}{n}
\newcommand{\rotate}{\mathbf{Q}}
\newcommand{\dipoleSusceptibility}{\boldsymbol{\susSym}_{\dipoleMag}}
\newcommand{\um}{\U_{\emag}}
\newcommand{\dsus}{\Delta \susSym}
\newcommand{\idenMat}{\mathbf{I}}
\newcommand{\mStateP}{\xvec}
\newcommand{\otherMStateP}{\mathbf{y}}
\newcommand{\reflAll}{\Sigma}
\newcommand{\tIn}{t_{\text{in}}}
\newcommand{\cStretch}{\lambda}
\newcommand{\pmag}{p}
\newcommand{\pvec}{\mathbf{\pmag}}